\documentclass[12pt]{article}

\usepackage{amsmath,array,amssymb,cite} % use drftcite in draftmode otherwise
\def\elabel#1{\label{#1}}

\DeclareSymbolFont{AMSb}{U}{msb}{m}{n}
\DeclareSymbolFontAlphabet{\Bbb}{AMSb}

\renewcommand{\thesection}{\Roman{section}}
\renewcommand{\theequation}{\arabic{section}.\arabic{equation}}
\def\rsen{\setcounter{equation}{0}}

\newcommand{\startappendix}{
\setcounter{section}{0}
\renewcommand{\thesection}{\Alph{section}}
\renewcommand{\theequation}{\Alph{section}.\arabic{equation}}}

\newcommand{\Appendix}[1]{
\refstepcounter{section}
\begin{flushleft}
{\large\bf Appendix \thesection: #1}
\end{flushleft}}

\newcommand{\aD}{{\dot\alpha}}

\def\Z{{\cal Z}}
\def\N{{\cal N}}
\def\det{{\rm det}}

\def\G{{G}}
\def\M{{\cal M}}
\def\tr{{\rm tr}}
\def\trtwo{\tr^{}_2\,}
\def\Mbar{\bar{\cal M}}
\def\dalpha{{\dot\alpha}}
\def\dbeta{{\dot\beta}}

\def\dmuphys{d\mu^{k}_{\rm phys}}

\def\sqrtwo{\sqrt{2}\,}
\def\hf{{\textstyle{1\over2}}}
\def\wbar{\bar w}
\def\mubar{\bar\mu}
\def\abar{\bar a}
\def\sigmabar{\bar\sigma}
\def\etabar{\bar\eta}
\def\zetabar{\bar\zeta}
\def\mubar{\bar\mu}
\def\nubar{\bar\nu}

\def\Tr{{\rm Tr}}

\def\dmuphys{d\mu^k_{\rm phys}}
\def\N{{\cal N}}

\def\P{{\cal P}}
\def\J{{\cal J}}
\def\G{{\cal G}}

\def\rmv{{\rm v}}

\def\sst{\scriptscriptstyle}

\def\Mbar{\bar{\M}}

\def\D{{\cal D}}
\def\Dbarslash{\,\,{\raise.15ex\hbox{/}\mkern-12mu {\bar\D}}}
\def\delslash{\,\,{\raise.15ex\hbox{/}\mkern-9mu \partial}}
\def\Dslash{\,\,{\raise.15ex\hbox{/}\mkern-12mu \D}}

\def\sigmabar{\bar\sigma}

\def\mubar{\bar\mu}
\def\dalpha{{\dot\alpha}}

\def\etabar{\bar\eta}

\def\Deltabar{\bar\Delta}

\def\N{{\cal N}}
\def\M{{\cal M}}

\def\hf{{\textstyle{1\over2}}}
\def\quarter{{\textstyle{1\over4}}}

\def\dbeta{{\dot\beta}}
\def\abar{\bar a}
\def\wbar{\bar w}
\def\trtwo{\tr^{}_2\,}
\def\trN{\tr^{}_N\,}

\def\nubar{\bar{\nu}}

\def\bbar{\bar b}

\def\sqrtwo{\sqrt{2}\,}

\def\CO{{\cal O}}

\def\hf{{\textstyle{1\over2}}}
\def\DRbar{\overline{\rm DR}}
\def\rmv{{\rm v}}
\def\rmvbar{\bar{\rm v}}
\def\rmvtilde{\tilde{\rm v}}
\def\trtwo{\tr^{}_2\,}
\def\det{{\rm det}}
\def\quarter{{\textstyle{1\over4}}}
\def\sqrtwo{\sqrt{2}\,}
\def\bbox{\square}
\def\G{{\cal G}}
\def\sumu{\sum_{u=0}^{N-1}}
\def\N{{\cal N}}
\def\M{{\cal M}}
\def\J{{\cal J}}
\def\P{{\cal P}}
\def\I{{\cal I}}

\def\rmP{{\Bbb P}}
\def\dalpha{{\dot\alpha}}
\def\aD{\dalpha}
\def\dbeta{{\dot\beta}}
\def\abar{\bar a}
\def\bbar{\bar b}
\def\nubar{\bar \nu}
\def\wbar{\bar w}
\def\mubar{\bar\mu}
\def\Mbar{\bar\M}
\def\zetabar{\bar\zeta}
\def\sigmabar{\bar\sigma}
\def\etabar{\bar\eta}
\def\Deltabar{\bar\Delta}
\def\bra#1{\left\langle #1\right|}
\def\ket#1{\left| #1\right\rangle}

\def\VEV#1{\left\langle #1\right\rangle}
\def\Vev#1{\big\langle{#1}\big\rangle}
\let\vev=\VEV
\def\kinst{{k\hbox{-}\rm inst}}
\def\trN{\tr_{N\,}^{}}
\def\trk{\tr_{k\,}^{}}

\begin{document}

\addtolength{\baselineskip}{4pt}
\thispagestyle{empty}

\begin{flushright}
{\tt hep-th/9904116}\\
April 1999
\end{flushright}

\vspace{1cm}

\begin{center}
{\scshape\Large Breakdown of Cluster Decomposition in\\ 
\vspace{0.15cm}
Instanton Calculations of the Gluino Condensate\\}

\vspace{1cm}

{\scshape Timothy J.~Hollowood$^{1,3}$,
Valentin V.~Khoze$^2$, Weonjong Lee$^1$\\
and Michael P.~Mattis$^1$}

\vspace{0.3cm}
$^1${\sl Theoretical Division T-8, Los Alamos National Laboratory,\\
Los Alamos, NM 87545, USA}\\

\vspace{0.2cm}
$^2${\sl Department of Physics, University of Durham,\\
Durham, DH1 3LE, UK}\\

\vspace{0.2cm}
$^3${\sl Department of Physics, University of Wales Swansea,\\
Swansea, SA2 8PP, UK}\\

\vspace{1cm}

{\Large ABSTRACT}
\end{center}

\vspace{0.1cm}

\noindent 
A longstanding puzzle concerns the calculation of the gluino
condensate $\Vev{\tr \lambda^2\over16\pi^2}=c\Lambda^3$ in $\N=1$
supersymmetric $SU(N)$ gauge theory: so-called weak-coupling instanton
(WCI) calculations give $c=1,$ whereas strong-coupling instanton (SCI)
calculations give, instead, $c=2\big((N-1)!(3N-1)\big)^{-1/N}.$ By
examining correlators of this condensate in arbitrary multi-instanton
sectors, we cast serious doubt on the SCI calculation of $\Vev{\tr
\lambda^2\over16\pi^2}$ by showing that an essential step---namely
cluster decomposition---is invalid. We also show that the addition of
a so-called Kovner-Shifman vacuum (in which  $\Vev{\tr
\lambda^2\over16\pi^2}=0$) cannot straightforwardly resolve this
mismatch.

\newpage

\section{Introduction}

Almost dating back to the development of QCD itself, supersymmetric
versions of QCD have been closely studied, as tractable laboratories
for extracting exact analytic information about both perturbative and
non-perturbative phenomena in nonabelian gauge theories.
One outstanding puzzle, unresolved since the mid-1980's, concerns the
calculation of the gluino condensate $\Vev{\tr \lambda^2\over16\pi^2}$
in these models. This is an interesting quantity, as it is a measure
of chiral symmetry breakdown. In pure supersymmetric Yang-Mills (SYM)
theory, by dimensional analysis, one expects
\begin{equation}
\VEV{\tr \lambda^2\over16\pi^2}\ =\ c\Lambda^3\ ,
\elabel{expect}\end{equation}
where $\Lambda$ is the dynamical scale in the theory (developed by
dimensional transmutation as in QCD), while $c$ is a numerical
constant. Remarkably, there are two  approaches in the
literature for calculating
$\Vev{\tr \lambda^2\over16\pi^2}$, each purporting to be exact (i.e.,
nonrenormalized), but which differ in their predictions of the
constant $c$. This disagreement is especially vexing in light of the
fact that both involve the use of supersymmetric instantons. The first
approach, generally known as ``strong-coupling instanton'' (SCI)
calculations, was developed in 
Refs.~\cite{Novikov:1983ee,Rossi:1984bu,Amati:1985uz,Amati:1988ft,Fuchs:1986ft}, 
while the second approach,
generally known as ``weak-coupling instanton'' (WCI) calculations, was
developed in Refs.~\cite{Fuchs:1986ft,Affleck:1983rr,Novikov:1985ic,Shifman:1988ia}; 
for self-containedness, both will be reviewed below. 

In this paper, we re-examine this old controversy, using our recently
developed methods for studying supersymmetric \it multi\rm-instantons
\cite{MO-I,MO-II,KMS,MO-III}. In particular, by looking at $n$-point
correlators
$\Vev{{\tr\lambda^2(x_1)\over16\pi^2}\cdots{\tr\lambda^2(x_n)\over16\pi^2}}$
of the gluino condensate, we will be able to probe arbitrary
topological numbers $k$. In a nutshell, our results cast serious doubt
on the validity of the SCI calculations of the
condensate. Specifically, we will demonstrate that an essential
technical step in the SCI approach, namely the use of cluster
decomposition, is invalid.  The important implications of this
observation are as follows. Since cluster decomposition is an
essential requirement of quantum field theories (with very mild
assumptions that are certainly met by SYM theory), the exact quantum
correlators must have this property. That cluster is violated by the
instanton-saturated SCI correlators then means that (contrary to
claims in the literature) the SCI approximation is only giving \it
part \rm of the full answer. Since the SCI correlators obey
supersymmetric perturbative nonrenormalization theorems
\cite{Novikov:1985ic}, it necessarily follows that additional
$non$-perturbative objects must be contributing to the correlators. A
fuller discussion of this point is given in Sec.~VII below; however,
categorizing the nature of these non-perturbative configurations is
beyond the scope of the present paper.\footnote{See however
Ref.~\cite{DHKM} where, in a compactified version of the present
theory, the important role played by monopoles is emphasized.}  We
should add that we believe that, in contrast, the WCI correlators are
consistent with cluster decomposition.

In addition, we will address an ingenious, if
controversial,\footnote{See Ref.~\cite{Csaki:1997aw} and the rebuttal
Ref.~\cite{Kogan:1998dt}.} hypothesis of Shifman's, in which the
numerical disagreement between the SCI and WCI results is taken as
circumstantial evidence for the existence of an extra disconnected
vacuum in SYM theory in which chiral symmetry is unbroken
\cite{Kovner:1997im,Shifman:1999mv}. While this so-called
``Kovner-Shifman (KS) vacuum'' can indeed potentially resolve the
disagreement at the 1-instanton sector ($k=1$), we will show that it
fails to do so for the topological sectors with $k>1.$ In other words,
positing a KS vacuum cannot by itself restore the cluster property to
the SCI correlators. This discouraging finding might be viewed as
removing some of the impetus for positing such a vacuum in the first
place.

Finally we will present a novel calculation of
$\Vev{\tr\lambda^2(x)\over16\pi^2}$ which relates the $\N=1$
supersymmetric models discussed herein to the exactly soluble
Seiberg-Witten models with $\N=2$ supersymmetry. This calculation is
of potential pedagogical interest because it bypasses the explicit use
of instantons, and instead relies on functional methods. Not
surprisingly, it recaptures the WCI answer. 

Let us sketch in broad strokes the main differences between the SCI
and WCI calculations (a more detailed review will follow). For $\N=1$
supersymmetric $SU(N)$ gauge theory with no matter, the leading
coefficient of the $\beta$-function is  $b_0=3N$, so that
$\Lambda^3$ goes like an ``$N^{\rm th}$ root'' of an instanton:
$\Lambda^3\propto\exp(-8\pi^2/g^2N).$ This means that a na\"\i ve
1-instanton calculation of $\Vev{\tr\lambda^2\over16\pi^2}\,$---in which
$\lambda$ is simply replaced by its ``classical value'' as an adjoint
fermion zero mode in the instanton background, and all the instanton
collective coordinates, both bosonic and fermionic, are integrated
over---fails; specifically it gives a zero answer, due to unsaturated
Grassmann integrations.  In order to perform a sensible 1-instanton
calculation of $\Vev{\tr\lambda^2\over16\pi^2}$, two alternative, and
necessarily more elaborate, approaches suggest themselves. In the SCI
approach, one calculates the $N$-point correlator of this
condensate, which scales like $\exp(-8\pi^2/g^2),$ and is indeed
nonzero at the 1-instanton level. Furthermore, by a Ward identity
reviewed in the Appendix, it
is independent of the $N$ space-time insertion points $x_i.$ After
performing the requisite collective coordinate integration, one finds:
\begin{equation}
\VEV{{\tr\lambda^2(x_1)\over16\pi^2}\cdots{\tr\lambda^2(x_N)\over16\pi^2}}\
=\ {2^N\over(N-1)!\,(3N-1)}\,\Lambda^{3N}\ .
\elabel{SCIans}\end{equation}
In order to extract $\Vev{\tr\lambda^2\over16\pi^2}$ from the correlator
\eqref{SCIans}, one then invokes  cluster decomposition: taking
$|x_i-x_j|\gg\mu^{-1}$ where $\mu$ is the mass gap in this theory, and
remembering the constancy of the correlator, one replaces the
left-hand side of Eq.~\eqref{SCIans} simply by
$\Vev{\tr\lambda^2\over16\pi^2}^N.$ The net result thus reads:
\begin{equation}
\VEV{\tr\lambda^2\over16\pi^2}\ =\ 
 {2\over\big((N-1)!\,(3N-1)\big)^{1/N}}\ \Lambda^{3}\,e^{2\pi
iu/N}\qquad\hbox{(SCI result)}\  ,
\elabel{SCIansb}\end{equation}
where $u=0,\ldots,N-1$ indexes the $N$ vacua $\ket{u}$
of the $SU(N)$ theory, and reflects the ambiguity in taking the
$N^{\rm th}$ root of unity. In retrospect (as argued in 
Refs.~\cite{Amati:1985uz,Amati:1988ft}),
the reason why the na\"\i ve calculation of 
$\Vev{\tr\lambda^2\over16\pi^2}$ gives zero is that these $N$ vacua
are being averaged over and the phases cancel.

In contrast, in the WCI approach, one modifies the pure gauge theory by adding
matter superfields in such a way that $\Vev{\tr\lambda^2\over16\pi^2}$
itself (rather than a higher-point function thereof) receives a
nonzero contribution at the 1-instanton level. Next, one decouples
these extraneous matter fields by giving them a mass $M$, and taking the joint
limit $M\rightarrow\infty$ and $\Lambda\rightarrow0$ in the
manner dictated by renormalization group (RG)
decoupling. Matching onto the effective low-energy theory without
matter gives:
\begin{equation}
\VEV{\tr\lambda^2\over16\pi^2}\ =\ 
\Lambda^{3}\qquad\hbox{(WCI result)}\  .
\elabel{WCIans}\end{equation}
Note that the RG decoupling procedure forces the low-energy theory
into one of the $N$ degenerate vacua $\ket{u}$, which by
convention we take to be the one with real phase. The nomenclature
``strong coupling'' versus ``weak coupling'' used to designate these
differing approaches refers to the fact that, in the former, as in
QCD, the only scale in the problem is the dynamical scale $\Lambda,$
whereas in the latter, the existence of VEVs $\rmv_i$ of the matter
superfields permit a standard semiclassical expansion when the
dimensionless ratios $\Lambda/\rmv_i$ are all small. (The holomorphic
properties of SYM theory then permit the analytic continuation of the
answer beyond this regime.)

As mentioned above, it is possible to reconcile the two calculations
\eqref{SCIansb} and \eqref{WCIans} by positing the existence of an extra vacuum
$\ket{S}$ in which the condensate vanishes \cite{Kovner:1997im}. 
Specifically, if $p$ and $1-p$ represent
the probability weights in the vacuum sector of the theory for the
standard vacua $\{\ket{u}\}$, and for $\ket{S}$, respectively, and if one
takes
\begin{equation}p\ =\ {2^N\over(N-1)!\,(3N-1)}\ ,
\elabel{Shifmana}\end{equation}
then both the 1-instanton results can be understood. Unfortunately,
the multi-instanton calculations presented below show that the
mismatch between the SCI and the WCI calculations becomes more severe
for higher topological number $k$, and apparently cannot be reconciled
in this way for $k>1.$

This paper is organized as follows. In Secs.~II and III, respectively,  we review the SCI
and WCI calculations of
$\Vev{\tr\lambda^2\over16\pi^2}$ for general gauge
group $SU(N)$. Also in Sec.~III we
present an alternate, non-instanton-based derivation of this
condensate, specific to the gauge group $SU(2)$, which starts from the
Seiberg-Witten solution of the $\N=2$ model \cite{Seiberg:1994rs}
and flows to the $\N=1$
model, recapturing the WCI result. In Sec.~IV we discuss  cluster
decomposition in more depth, and motivate Shifman's proposal for
reconciling the SCI and WCI calculations by postulating an extra
vacuum state. Our principal results are described in Secs.~V and VI, in which
(extending the SCI approach) we calculate higher-point functions of
the condensate, in the topological sectors  $k>1.$ In
Sec.~V we calculate, analytically, the $(kN)$-point functions
$\Vev{{\tr\lambda^2(x_1)\over16\pi^2}\cdots{\tr\lambda^2(x_{kN})\over16\pi^2}}$
in $SU(N)$ gauge theory for arbitrary instanton number $k$, but to
leading order in $1/N,$ while in Sec.~VI we calculate, numerically, the
4-point function 
$\Vev{{\tr\lambda^2(x_1)\over16\pi^2}{\tr\lambda^2(x_2)\over16\pi^2}
{\tr\lambda^2(x_3)\over16\pi^2}   
{\tr\lambda^2(x_{4})\over16\pi^2}}$ at the 2-instanton level for
gauge group $SU(2)$. In either case our SCI calculations explicitly contradict
the hypothesis of cluster decomposition---both with and without an
extra KS vacuum.\footnote{The numerical calculation is based on a
Monte Carlo integration, which (with our present statistics) is
incompatible with the clustering result at the $5\hf$ sigma level,
and incompatible with the modified clustering result due to the
incorporation of a KS vacuum (tuned to reconcile the WCI and SCI
1-instanton results), at the $11\hf$ sigma level.}
 Concluding comments are made in Sec.~VII.

\rsen
\section{Review of the Strong-Coupling Instanton Calculation}

Let us review the SCI result for
$\Vev{\tr\lambda^2(x_1)\cdots\tr\lambda^2(x_N)}$, for pure $\N=1$
$SU(N)$ gauge theory. The calculation for done originally for the
$SU(2)$ theory in \cite{Novikov:1983ee} and then extended to the
$SU(N)$ theories in \cite{Amati:1985uz} (see also the very comprehensive
review articles \cite{Amati:1988ft,Shifman:1999mv}).

The correlator in question is saturated at the 1-instanton level. 
The gauge-invariant collective coordinate integration measure is a
suitable generalization of the Bernard measure \cite{Bernard} to an
${\cal N}=1$ theory,
and reads:\footnote{Our choice of notation is dictated by the
$k$-instanton generalization of this measure, Eq.~\eqref{GImeas}
below. Following Ref.~\cite{Finnell:1995dr}, we correct a factor of two
mistake in the normalization of adjoint fermion zero modes that
pervades much of the literature (e.g.,
Refs.~\cite{Amati:1988ft,Fuchs:1986ft}). Hence our final result for
the $N$-point function, Eq.~\eqref{SCIans}, differs by $2^N$ from
these references.}
\begin{equation}
-{2^{3N+2}\,\pi^{2N-2}\Lambda^{3N}\over(N-1)!\,(N-2)!}\int
d^4a'\,d\rho^2\,(\rho^2)^{2N-4}\,d^2\M'\,
d^2\zeta\,d^{N-2}\nu\,d^{N-2}\nubar\ .
\elabel{measone}\end{equation}
Here $a'_n$ is (minus) the
4-position of the instanton and $\rho$ is its scale size, the
Grassmann spinors $\M'_\alpha$ and $\zeta_\dalpha$ parametrize the
supersymmetric and superconformal modes, respectively, of the gluino,
and the Grassmann parameters $\nu_{u'}$ and $\nubar_{u'}$,
$u'=1,\ldots,N-2,$ are the superpartners to the iso-orientation modes
which sweep the instanton through $SU(2)$ subgroups of the $SU(N)$
gauge group (note that each $\nu_{u'}$ and $\nubar_{u'}$ is a
Grassmann number rather than a Grassmann spinor). The measure includes
the Lambda parameter of the Pauli-Villars (PV) scheme which at the two-loop
level is \cite{Amati:1988ft}
\begin{equation}
\Lambda=g(\mu)^{-2/3}e^{-8\pi^2/(3Ng(\mu)^2)}\mu\ .
\end{equation}
Into this measure
one inserts $\prod_{i=1}^N\tr\lambda^2(x_i)$ where $\lambda^\alpha(x)$
is the most general classical adjoint fermion zero mode in the
1-instanton background. In terms of these bosonic and fermionic
collective coordinates, one derives (see Eq.~\eqref{corriganid} below):
\begin{equation}\begin{split}\tr\lambda^2(x)\ =\ -
\quarter\bbox\Big(&{2\over\rho^2+y^2}\sum_{u'=1}^{N-2}
\nubar_{u'}\nu_{u'}\ +\ \zeta_\dalpha\zeta^\dalpha\,{y^4\over(\rho^2+y^2)^2}
\\&+\ \M^{\prime\alpha}\M'_\alpha\,{2\rho^2+y^2\over(\rho^2+y^2)^2}
\ -\ \M^{\prime\alpha}y_{\alpha\dalpha}\zeta^\dalpha\,{2\rho^2\over(\rho^2+y^2)^2}\Big)
\elabel{insertdef}\end{split}\end{equation}
where
\begin{equation}
y_{\alpha\dalpha}
\ =\ x_{\alpha\dalpha}+a'_{\alpha\dalpha}\ =\
(x_n+a'_n)\sigma^n_{\alpha\dalpha}\ .
\elabel{ydef}\end{equation}

Now let us carry out the Grassmann integrations in Eq.~\eqref{measone}.
Obviously the $\zeta$ and $\M'$ Grassmann integrations will be
saturated from the condensates inserted at two points
$\{x_i,x_j\}$ chosen from among the $N$ insertions
$x_1,\ldots,x_N$. For each such pair there are three contributions to
these integrals:
\def\sixteenth{{\textstyle{1\over16}}}
\begin{subequations}
\begin{align}
&\sixteenth\bbox_i\bbox_j\,{y_i^4(2\rho^2+y_j^2)\over
(\rho^2+y_i^2)^2(\rho^2+y_j^2)^2}\qquad(\zeta^2\hbox{ at } x_i,\
\M^{\prime2}\hbox{ at }x_j)\ ,\elabel{threecona}\\
&\sixteenth\bbox_i\bbox_j\,{(2\rho^2+y_i^2)y_j^4\over
(\rho^2+y_i^2)^2(\rho^2+y_j^2)^2}\qquad(\M^{\prime2}\hbox{ at } x_i,\
\zeta^{2}\hbox{ at }x_j)\ ,\elabel{threeconb}\\
&\sixteenth\bbox_i\bbox_j\,{2\rho^4\,y_i\cdot y_j\over
(\rho^2+y_i^2)^2(\rho^2+y_j^2)^2}\qquad(\zeta\times\M'\hbox{ at } x_i,\
\zeta\times\M^{\prime}\hbox{ at }x_j)\ .\elabel{threeconc}
\end{align}
\end{subequations}
Adding these three contributions gives the simpler expression
\begin{equation}
{-36\rho^8\,(x_i-x_j)^2\over (\rho^2+y_i^2)^4(\rho^2+y_j^2)^4}\ .
\elabel{adding}\end{equation} Now we take advantage of the fact that
this $N$-point function is independent of the $x_i$ (see the
Appendix), to choose these insertion points for maximum simplicity of
the algebra. The simplest conceivable such choice, $x_i=0$ for all
$i,$ turns out to give an ill-defined answer of the form
``$0\times\infty$'' (the zero coming from the Grassmann integrations
as follows from Eq.~\eqref{adding}, and the infinity from divergences
in the $\rho^2$ integration due to coincident poles).  In order to
sidestep this ambiguity, one chooses instead:
\begin{equation}
x_1=\cdots=x_{N-1}=0\ ,\qquad x_N=x\ .
\elabel{insertionone}\end{equation} This choice is the simplest one
which gives a well-defined answer with no ``$0\times\infty$''
ambiguity. More ambitiously, one can still perform the calculation
even if all the insertion points are taken to be arbitrary
\cite{Amati:1985uz,Amati:1988ft}; however, we find it convenient for
later to take the minimal resolution provided by
\eqref{insertionone}. From the $(x_i-x_j)^2$ dependence in
Eq.~\eqref{adding}, it follows that the pair of insertions
$\{x_i,x_j\}$ responsible for the $\{\zeta,\M'\}$ integrations must
include the point $x_N=x$; there are $N-1$ possible such pairs, giving
\begin{equation}
{-36(N-1)\rho^8\,x^2\over
\big(\rho^2+(x+a')^2\big)^4\,(\rho^2+a^{\prime2})_{}^4}
\elabel{addingb}\end{equation}
for these contributions. The remaining Grassmann integrations over
$\{\nu,\nubar\}$ are saturated at $x_i=0,$ and give
\begin{equation}
(N-2)!\,\Big({4\rho^2\over
(\rho^2+a^{\prime2})^3}\Big)^{N-2}\ .
\elabel{nuresult}\end{equation}

Combining the denominators in Eqs.~\eqref{addingb}-\eqref{nuresult} with a Feynman
parameter $\alpha$,
\begin{equation}
{1\over(\rho^2+a^{\prime2})^{3N-2}}\,
{1\over(\rho^2+(x+a^{\prime})^2)^4}\ =\
{(3N+1)!\over3!\,(3N-3)!}\int_0^1d\alpha\,{\alpha^3(1-\alpha)^{3N-3}\over\big(
\rho^2+(a'+\alpha x)^2+\alpha(1-\alpha)x^2\big)^{3N+2}}
\elabel{alphadef}\end{equation}
and performing the $d^4a'$ integration then yields:
\begin{equation}\begin{split}
\VEV{\tr\lambda^2(x_1)\cdots\tr\lambda^2(x_N)}\ &=\ 
{2^{3N+2}\,\pi^{2N-2}\Lambda^{3N}\over(N-1)!\,(N-2)!}\int_0^1d\alpha
\int_0^\infty d\rho^2\,(\rho^2)^{2N-4}
\\&\quad\quad\times\
\big(36(N-1)\rho^8x^2\big)(N-2)!\,(4\rho^2)^{N-2}
\\&\quad\quad\times\ 
{(3N+1)!\over3!\,(3N-3)!}\,{\alpha^3(1-\alpha)^{3N-3}\,\pi^2\over3N(3N+1)\big(
\rho^2+\alpha(1-\alpha)x^2\big)^{3N}}\\&=\
{3(3N-2)\,2^{5N-1}\,\pi^{2N}\Lambda^{3N}\over(N-2)!}\int_0^1d\alpha\,\alpha^2(1-\alpha)^{3N-4}
\\&=\ {2^{5N}\,\pi^{2N}\Lambda^{3N}\over(N-1)!\,(3N-1)}\ ,
\elabel{finalone}\end{split}\end{equation}
in agreement with Eqs.~\eqref{SCIans}-\eqref{SCIansb}.

\rsen
\section{Review of the Weak-Coupling Instanton Calculation}

Next, let us review the WCI calculation of the gluino condensate. As
mentioned above, the general WCI strategy is to extend the pure gauge
theory to include matter content, in such a way that
$\Vev{\tr\lambda^2\over16\pi^2}$ receives a nonzero contribution at
the 1-instanton level. Decoupling the extraneous matter and matching
to the low-energy pure gauge theory is then accomplished using
standard RG prescriptions. Since the precise nature of this extraneous
matter is rather arbitrary, the WCI calculation really stands for a
family of related calculations sharing this basic approach, all of
which give the same result \eqref{WCIans}. Calculations of this type
were done in
\cite{Affleck:1983rr,Novikov:1985ic,Fuchs:1986ft,Shifman:1988ia} 
and reviewed in \cite{Shifman:1999mv}.

We will find it efficient to exploit the functional identity (see for
example \cite{Peskin:1997qi}):
\def\Weff{{\cal W}_{\rm eff}}
\def\ddtau{{\partial\over\partial\tau}}
\def\ddLambda{{\partial\over\partial\Lambda}}
\begin{equation}
\VEV{\tr\lambda^2}\ =\ -8\pi i\VEV{\ddtau\Weff}\ =\
{16\pi^2\over b_0}\VEV{\Lambda\ddLambda\Weff}\ .
\elabel{fcnlid}\end{equation}
Here $\Weff$ is the effective superpotential,
\begin{equation}
\tau\ =\ {4\pi i\over g^2}+{\theta\over2\pi}
\elabel{taudef}\end{equation}
is the usual complexified coupling, and
\begin{equation}
\Lambda\ =\ \mu\,e^{2\pi i\tau(\mu)/b_0}
\elabel{Lambdadef}\end{equation}
is the RG-invariant 1-loop dynamical scale of the theory. This result comes
from writing the microscopic gauge theory as
\def\L{{\cal L}}
\def\Im{{\rm Im}}
\begin{equation}
\L\ =\ {1\over4\pi}\,\Im\big(\tau\int
d^2\theta\,\tr\,W^{\alpha}W_\alpha\big)\ ,
\elabel{microgt}\end{equation}
where $W^\alpha$ is the gauge field-strength chiral superfield,
and promoting $\tau$ to a ``spurion superfield'',
\def\Wadef{W^a\ =\ \lambda^a+\cdots}
\begin{equation}
\tau\ \rightarrow\ T(y,\theta)\ =\
\tau(y)+\sqrtwo\theta^\alpha\chi^\tau_\alpha(y) +\theta^2F^\tau(y)\ .
\elabel{taupromote}\end{equation}
{}From Eqs.~\eqref{microgt}-\eqref{taupromote} it trivially follows that
\def\Z{{\cal Z}}
\def\condition{\,{\Big|}_{T(y,\theta)=\tau}}
\begin{equation}
\VEV{\tr\lambda^2}\ =\ {8\pi\over\Z}{\delta\over\delta
F^\tau(x)}\,\Z\condition\ ,
\elabel{Zone}\end{equation}
where
\def\D{{\cal D}}
\begin{equation}\Z\ =\ \int\D W\,e^{i\int d^4x\,\L}
\elabel{Ztwo}\end{equation} 
is the partition function of the
microscopic theory, in the generalized background field
\eqref{taupromote}. In order to derive Eq.~\eqref{fcnlid} from
Eq.~\eqref{Zone}, one assumes that the functional differentiation
indicated in Eq.~\eqref{Zone} formally commutes with the
integrating-out of the microscopic degrees of freedom. In other words,
$\Z$ can be re-expressed in terms of the relevant effective chiral
superfields $\Phi_i$ (whatever these may be\footnote{In the example
culminating in Eq.~\eqref{adsdefc} below, we will find that there are
in fact no residual chiral superfields `$\Phi_i$', so that simply
$\Weff=\Weff(T),$ whereas in the Seiberg-Witten example \eqref{WSWdef}
below, the $\Phi_i$ are the monopole superfields $M$, $\tilde M$ as
well as the dual Higgs $A_D$.}): 
\def\Zeff{\Z_{\rm eff}}
\begin{equation}\VEV{\tr\lambda^2}\ =\  {8\pi
\over\Zeff}{\delta\over\delta F^\tau(x)}\,\Zeff\condition\ ,
\elabel{Zthree}\end{equation}
where 
\begin{equation}
\Zeff\ =\ \int\D\Phi_i\,e^{-i\int d^4x\int
d^2\theta\,\Weff(\Phi_i,T)}\ .
\elabel{Zfour}\end{equation}
Equation \eqref{fcnlid} then follows from the observation that
$\partial\Weff/\partial F^\tau=\theta^2\,\partial\Weff/\partial\tau$.

We now need an explicit expression for the effective superpotential.
Following Affleck, Dine and Seiberg (ADS) \cite{Affleck:1983rr}, 
it is convenient to start
from $SU(N)$ gauge theory where the number of flavors $N_F$ is fixed
to $N_F=N-1.$ A 1-instanton calculation of the superpotential then
gives:
\def\cads{C_{\rm \scriptscriptstyle ADS}}
\def\Qtilde{{\tilde Q}}
\begin{equation}\Weff^{N_F,N}\ \equiv\ \Weff^{N-1,N}\ =\
\cads\,{\Lambda^{b_0}_{N-1,N}\over \det_{N_F}\big(Q_f\Qtilde_{f'}\big)}\ ,
\elabel{adsdef}\end{equation}
where the flavor indices $f,f'=1,\ldots,N_F$ run over the quark
superfields. The coefficient of the $\beta$-function is, for general
$N$ and $N_F,$
\begin{equation}b_0\ =\ 3N-N_F\ .
\elabel{bzerodef}\end{equation}
The normalization constant for the specific case $N_F=N-1$ was fixed
by an explicit 1-instanton calculation, and is simply
\cite{Cordes,Finnell:1995dr} $\cads=1$. By 
decoupling the quark flavors one at a time, this 1-instanton
expression flows into models with $N_F<N-1$ for which the
superpotential is no longer a 1-instanton phenomenon. In this way one
generalizes Eq.~\eqref{adsdef} to (see e.g.,
Refs.~\cite{Peskin:1997qi,Intriligator:1996au}):
\begin{equation}
\Weff^{N_F,N}\  =\
\cads^{N_F,N}\,\left({\Lambda^{b_0}_{N_F,N}\over
\det_{N_F}\big(Q_f\Qtilde_{f'}\big)}\right)^{1\over N-N_F}\quad(N_F\le
N-1)\ ,
\elabel{adsdefb}\end{equation}
where 
\cite{Finnell:1995dr}
\begin{equation}\cads^{N_F,N}\ =\ N-N_F\ .
\elabel{cadsdefa}\end{equation}

Starting from this more general superpotential, let us decouple the
remaining quarks, by giving them a common VEV v. Viewing $Q$ as an
$N_F\times N$ matrix, one assumes:
\begin{equation}
\VEV{Q}\ =\
\begin{pmatrix}\rmv&{}&0&0&\cdots&0\\{}&\ddots&{}&\vdots&{}&\vdots\\ 
0&{}&\rmv&0&\cdots&0\end{pmatrix}\ ,\quad 
\langle\Qtilde\rangle\ =\
\begin{pmatrix}\rmvtilde&{}&0\\{}&\ddots&{}\\0&{}&\rmvtilde\\0&\cdots&0\\\vdots&{}&\vdots\\
0&\cdots&0\end{pmatrix}\ .
\elabel{Qvev}\end{equation}
The $D$-flatness condition together with a global gauge rotation gives $\rmvtilde=\rmvbar.$
Taking $|\rmv|\rightarrow\infty$ then decouples the quarks as well as a
subset of the gauge fields, leaving a pure $SU(N')$ gauge theory with
$N'=N-N_F$  and $b_0=3N'$. The 1-loop RG matching prescription reads
\cite{Finnell:1995dr}:
\begin{equation}
\left({\Lambda_{N_F,N}\over|\rmv|}\right)^{3N-N_F}\ =\ 
\left({\Lambda_{0,N'}\over|\rmv|}\right)^{3N'}\ ,\quad N'=N-N_F\ .
\elabel{matching}\end{equation}
Inputting Eqs.~\eqref{cadsdefa}-\eqref{matching} into Eq.~\eqref{adsdefb} gives:
\begin{equation}\Weff\  =\
(N-N_F)\,\left({\Lambda^{3N-N_F}_{N_F,N}\over|\rmv|^{2N_F}}
\right)^{1\over N-N_F}\ =\ N'\big(\Lambda_{0,N'}\big)^3\ .
\elabel{adsdefc}\end{equation}
The desired result \eqref{WCIans} then follows from Eq.~\eqref{fcnlid}.

Note that the starting-point for this WCI calculation, Eq.~\eqref{adsdef}, is
a \it bona fide \rm \hbox{1-instanton} calculation. The remaining steps
towards the answer involve well-studied path-integral and renormalization group
manipulations (principally Eq.~\eqref{fcnlid}, and Eqs.~\eqref{adsdefb}-\eqref{matching},
respectively). Alternatively, starting again from the functional
identity \eqref{fcnlid}, we can rederive the WCI result \eqref{WCIans} without
any reference to an instanton calculation. Instead, one starts from the
Seiberg-Witten solution of the $\N=2$ model,\footnote{For the remainder of
the section, we focus on $SU(2)$ gauge theory, and quote well-known
formulae from Seiberg and Witten \cite{Seiberg:1994rs}.} 
in the presence of a mass
deformation which breaks the supersymmetry down to $\N=1.$ In the
strong-coupling domain, in the vicinity of the monopole singularity,
the superpotential looks like:
\def\WSW{{\cal W}_{\scriptscriptstyle\rm SW}}
\begin{equation}
\WSW\ =\ \sqrtwo\tilde MA_DM\,+\,m\,U(A_D)\ .
\elabel{WSWdef}\end{equation}
Here the chiral superfields $\{M,\tilde M\}$ describe the monopole
multiplet, $A_D$ is the dual Higgs, $U$ is the quantum modulus of the
theory (here, in strong coupling,  expressed in terms of $A_D$ rather
than $A$), and $m$ is the mass parameter. The $F$-flatness condition
for the vacuum reads
\begin{equation}
0\ =\ {\partial\WSW\over\partial M}\ =\
{\partial\WSW\over\partial\tilde M}\ =\ {\partial\WSW\over\partial
A_D}\ ,
\elabel{Fflat}\end{equation}
which is solved by
\begin{equation}
a_D\equiv \VEV{A_D}=0\ ,\qquad \VEV{M}=\langle\tilde M\rangle=
\Big(-{m\over\sqrtwo}U'(0)\Big)^{1/2}\ .
\elabel{Fflatsolve}\end{equation}
In the vicinity of this solution, the relationship between $a_D$ and
$u=\Vev{U}$ is given by
\def\LamSW{\Lambda_{\scriptscriptstyle\rm SW}}
\def\Lambdatwo{\Lambda_{\scriptscriptstyle\N=2}}
\begin{equation}
a_D\ =\ {\sqrtwo\over\pi}\int_1^udx\,{\sqrt{x-u}\over\sqrt{x^2-\LamSW^4}}
\elabel{aDdef}\end{equation}
from which it follows that
\begin{equation}
u\ =\ \LamSW^2-2i\LamSW a_D+\CO(a_D^2)\ .
\elabel{useries}\end{equation}
Here the Seiberg-Witten dynamical scale $\LamSW$ is related to the
conventional PV/$\DRbar$ scale $\Lambdatwo$ via \cite{Finnell:1995dr}
\begin{equation}
\LamSW\ = \sqrtwo\Lambdatwo\ .
\elabel{LamSWdef}\end{equation}
Note that the series \eqref{useries} is not an instanton expansion (i.e., an
expansion in $\LamSW^4$); instantons emerge only in
the weak-coupling regime, where $u$ is expanded in terms  of $a=\Vev{A}$ rather
than $a_D$. 

Applying the identity \eqref{fcnlid} to $\WSW$ using
Eqs.~\eqref{useries}-\eqref{LamSWdef} gives the gluino condensate in the vacuum
\eqref{Fflatsolve}:
\begin{equation}
\VEV{\tr \lambda^2}\ =\ 16\pi^2m\Lambdatwo^2\ .
\elabel{SWans}\end{equation}
Next we decouple the adjoint Higgs superfield, by taking
$m\rightarrow\infty.$ In this way we flow to the pure $\N=1$
supersymmetric $SU(2)$ gauge theory.
The RG matching condition between the scale $\Lambda$ of the $\N=1$ theory and the
scale $\Lambdatwo$ of the mass-deformed $\N=2$ theory reads \cite{Finnell:1995dr}:
\begin{equation}
m^2\Lambdatwo^4\ =\ \Lambda^6\ .
\elabel{SWmatch}\end{equation}
Substituting Eq.~\eqref{SWmatch} into Eq.~\eqref{SWans} once again gives the WCI
answer \eqref{WCIans}.

\rsen
\section{Comments on Cluster Decomposition}

In this section, we examine the issue of cluster decomposition in the
context of the gluino condensate. This issue of cluster decomposition
is fundamental to a quantum field theory.  The clustering property
requires that for sufficiently large separations $|x_i-x_j|$, compared
with the inverse mass gap,\footnote{For a discussion of clustering and
other references, see Bogolubov {\it et al.\/} \cite{BOG}.}
\begin{equation} 
\VEV{\varphi_1(x_1)\cdots\varphi_n(x_n)}\ \rightarrow\ 
\VEV{\varphi_1}\times\cdots\times\VEV{\varphi_n}\ .
\end{equation}
Generally, this property breaks down when, in a statistical mechanical
sense, the theory is in a mixed phase. In field theory language, this
means there is more than one possible vacuum state. The clustering property
is then restored by restricting the theory to the Hilbert space built
on one of the vacua. In this sense, clustering is violated in a mild
way, and to distinguish this from some other, potentially more serious, 
violations uncovered below, we will
say that the theory satisfies a ``generalized notion of clustering''.
 
Let us consider the calculation of the $\G_n$, the $n$-point function
of the composite operator $\trN\lambda^2\,$:
\begin{equation}\G_n(x_1,\ldots,x_n)\ =\ \langle\trN\lambda^2(x_1)\cdots
\trN\lambda^2(x_n)\rangle\ .
\elabel{Gndef}\end{equation}
For present purposes we restrict our attention to pure $\N=1$ $SU(N)$
gauge theory.  Since $\trN\lambda^2$ is the lowest component of a
gauge-invariant chiral superfield (namely $\trN W^2$ where $W^\alpha$
is the field-strength superfield), a well-known identity---reviewed in
the Appendix---says that
\begin{equation}
\G_n(x_1,\ldots,x_n)\ = \hbox{const.\ ,}
\elabel{Gnconst}\end{equation}
independent of the $x_i$. Next let us consider this constant
correlator in the instanton
approximation. This means
that, at topological level $k$, $\lambda(x)$ is simply to be replaced
by a general superposition of adjoint fermion zero modes in the
general ADHM $k$-instanton background, weighted by Grassmann-valued
parameters (i.e., fermionic collective coordinates).  All bosonic and
fermionic collective coordinates are then integrated over, in the
appropriate supersymmetric way reviewed below. It can also be shown
that $\G_n$ should still be a constant. (The field theory proof of the
constancy of the correlation functions and its extension to the
instanton approximation is discussed in the Appendix.) Now, in $SU(N)$
gauge theory, at the topological level $k$, a multi-instanton has
precisely $2kN$ adjoint fermion zero modes which need to be integrated
over. Let us summarize the rules
for Grassmann integration: if $\xi$ is a Grassmann parameter, then
\begin{equation}
\int d\xi\,\xi=1\ ,\quad\int d\xi\,1=0\ .
\elabel{xiint}\end{equation}
Since $\trN\lambda^2$ is a Grassmann bilinear, it follows that $\G_n$
is only non-vanishing for $n=kN$. In particular, the one-point function
$\G_1$ always vanishes.  In summary, in the instanton
approximation, at topological level $k$, we have the following
selection rule:
\begin{subequations}
\begin{align}
&\langle\trN\lambda^2(x)\rangle\Big|_\kinst\ 
\equiv \ \G_1\,\Big|_\kinst\ 
=\ 0\quad\hbox{for all}\ k\ ;\elabel{selrulesa}\\
&\langle\trN\lambda^2(x_1)\cdots
\trN\lambda^2(x_n)\rangle\Big|_\kinst\ \equiv\ \G_n\Big|_\kinst\ \neq\
0 \quad\hbox{if and only if}\ n=kN\ .\elabel{selrulesb}
\end{align}
\end{subequations}
Notice that these results already indicate a breakdown of clustering
for the correlation functions \eqref{Gndef}, although, as we shall
explain below the breakdown is of the 
`mild' variety and can be traced to the fact that in
instanton approximation the theory is in a mixed phase, i.e.~the
instanton approximation samples the theory in a number of distinct
vacua as opposed to a single vacuum.

A general field-theoretic understanding of the selection rule
\eqref{selrulesa}-\eqref{selrulesb} was 
suggested in Refs.~\cite{Amati:1985uz,Amati:1988ft}. The suggestion
%does not rely on the mathematical construct of Grassmann integration, 
relies on the fact 
that, in $\N=1$ $SU(N)$ gauge theory, the vacua of the theory come in
an $N$-tuplet \cite{Witten:1982df}. The vacua spontaneously break the
discrete ${\Bbb Z}_{2N}$ anomaly-free remnant of the classical
$U(1)_R$ symmetry to the 
${\Bbb Z}_2$ subgroup: $\lambda_\alpha\rightarrow-\lambda_\alpha$. 
The vacuum sector therefore consists of:
\begin{equation}\big\{\ \ket{u}\ :\quad0\le u\le N-1\ \big\}\ .
\elabel{ketjdef}\end{equation}
If we define the condensate $\J$ via
\begin{equation}
\J\ =\ \bra{u=0}\trN\lambda^2\ket{u=0}\ ,
\elabel{Jdef}\end{equation}
then the $N$-tuple of vacua are related by phase factors, namely the
$N^{\rm th}$ roots of unity:
\begin{equation}
\bra{u}\trN\lambda^2\ket{u}\ =\ \J\,e^{2\pi iu/N}\ .
\elabel{jvevs}\end{equation}
Now let us see how the selection rule \eqref{selrulesa} comes about. We
define the density matrix
\begin{equation}
\varrho\ =\ {1\over N}\sum_{u=0}^{N-1}\ket{u}\bra{u}\ .
\elabel{densmat}\end{equation}
Since the instanton calculation is ${\Bbb Z}_{2N}$ symmetric, it must
{\it average\/} over all the vacua. This means
\begin{equation}\vev{\trN\lambda^2}=\Tr\big(\varrho\,\trN\lambda^2\big)
={1\over N}\sum_{u=0}^{N-1}\bra{u}\trN\lambda^2\ket{u}={\J\over N}
\sumu e^{2\pi iu/N}=0\ .
\elabel{firstsel}\end{equation}
Here the capitalized `$\Tr$' means a trace over the Hilbert space. In
order to check the selection rule \eqref{selrulesb}, we need the additional
assumption of a well-defined clustering limit. 
We have:
\begin{equation}\begin{split}
&\langle\trN\lambda^2(x_1)\cdots
\trN\lambda^2(x_n)
\rangle\ =\ \Tr\big(\varrho\,
\trN\lambda^2(x_1)\cdots
\trN\lambda^2(x_n)\big)\\ &\qquad\qquad\qquad=\
{1\over N}\sumu\bra{u}\trN\lambda^2(x_1)\,\rmP\,\trN\lambda^2(x_2)\,\rmP\cdots\rmP\,
\trN\lambda^2(x_n)\ket{u}\ ,
\elabel{npoint}\end{split}\end{equation}
where $\rmP$ denotes the sum over a complete set of states. At this
point the generalized notion of the clustering assumption enters. 
We assume there exists a mass
gap $\mu$ that is dynamically generated in the theory, and we consider
the $n$ insertion points are sufficiently far separated in Euclidean
space compared to this scale: $|x_i-x_j|\gg\mu^{-1}.$ (Since $\G_n$ is
a constant even in leading semiclassical order, moving to this regime
does not entail any additional approximations.) 
In this regime, the generalized cluster decomposition
(in our present usage) is equivalent
to the statement that $\rmP$ collapses to $\rmP_0$ where $\rmP_0$ is the
projection operator onto vacuum states only:
\begin{equation}
\rmP\rightarrow\rmP_0\ ,\qquad
\rmP_0\ =\ \sumu\ket{u}\bra{u}\ .
\elabel{Pzerodef}\end{equation}
Using the fact that the operator $\trN\lambda^2$ is diagonal in the
$u$ index,\footnote{This follows from the fact that there should be no mixing
between the sectors of Hilbert space built on each vacuum: they are
{\it super-selection sectors\/}.} it follows that with the replacement \eqref{Pzerodef},
the correlator \eqref{npoint} collapses to
\begin{equation}
{1\over N}\sumu\big(\J\,e^{2\pi i u/N}\big)^n\ =\
\begin{cases}
\J^{kN}\ & n=kN\\ 0\ & \hbox{otherwise}\ .\end{cases}
\elabel{collapse}\end{equation}

Next we consider how this elementary analysis is modified if the
$N$-tuplet of vacua $\{\ket{u}\}$ is supplemented by an extra vacuum
state, the so-called Kovner-Shifman vacuum \cite{Kovner:1997im}, which we denote
$\ket{S}$. A single such vacuum is permissible under the discrete symmetry only if
\begin{equation}
\bra{S}\trN\lambda^2\ket{S}\ =\ 0\ .
\elabel{KScond}\end{equation}
The analysis proceeds just as before, with the obvious modification
that the density matrix $\varrho$ should be replaced by $\varrho',$
defined by
\begin{equation}
\varrho'\ =\ (1-p)\ket{S}\bra{S}\ +\ {p\over N}\sumu\ket{u}\bra{u}\ ,
\elabel{varrhopdef}\end{equation}
where  the probability $p$ is a real number between 0 and 1. 
Proceeding as before, we find that with the generalized clustering
assumption 
\begin{equation}
\rmP\rightarrow\rmP'_0\ ,\qquad
\rmP'_0\ =\ \ket{S}\bra{S}\ +\ \sumu\ket{u}\bra{u}\ ,
\elabel{Pzerodefa}\end{equation}
one derives
\begin{equation}
\langle\trN\lambda^2(x_1)\cdots\trN\lambda^2(x_n)\rangle\ =\ 
\begin{cases}p\J^{kN}\ & n=kN\\ 0\ & \hbox{otherwise}\ .\end{cases}
\elabel{collapseb}\end{equation}
Obviously this modified expression also applies if there are several
distinct KS vacua $\ket{S_i}$ in which $\tr\lambda^2=0,$ that is 
\begin{equation}
\varrho'\ =\ \sum_{i=1}^l q_i\ket{S_i}\bra{S_i}\ +\ {p\over N}\sumu\ket{u}\bra{u}\ ,
\elabel{varrhopdefz}\end{equation}
where $p=1-\sum q_i$.

In the following we will calculate these $(kN)$-point correlators,
first analytically for large $N$, then numerically for $N=2$ and $k=2$, and will
find a behavior quite different from either \eqref{collapse} or \eqref{collapseb}.

\rsen
\section{Large-$N$ Calculation of Gluino Condensate Correlation Functions}

We now present an explicit evaluation of $\G_n,$ $n=kN,$ in the limit
$N\rightarrow\infty$ with $k$ held fixed. Our answer turns out to be
incompatible with both Eqs.~\eqref{collapse} and \eqref{collapseb}. The cleanest
way to quantify this disagreement is to consider the $(kN)^{\rm th}$ root,
$(\G_{kN})^{1/kN}.$ In the large-$N$ limit, from
Eq.~\eqref{collapse}, i.e.~clustering without the KS vacuum,
%or Eq.~\collapseb, 
one obtains:
\begin{equation}
\lim_{N\rightarrow\infty}(\G_{kN})^{1/kN}\ =\ \J(N)\ .
\elabel{rootdef}\end{equation}
We have written $\J$ as $\J(N)$ to allow for an unknown $N$
dependence. Using the one instanton expression \eqref{SCIansb} in the
large-$N$ limit one expects 
\begin{equation}
{1\over16\pi^2}{\cal J}(N)={2e\over N}\,\Lambda^3\ ,
\elabel{jnclust}\end{equation}
where $e=2.718\cdots.$
The key point is that the right-hand side of \eqref{rootdef} is
independent of the topological number $k$ (as well as of the
space-time insertion points $x_i$). Note that Eq.~\eqref{rootdef} follows,
not only from Eq.~\eqref{collapse}, but also
from Eq.~\eqref{collapseb}, so long as the constant $p$ either has a nonzero
large-$N$ limit, or else vanishes at large $N$
more  slowly than exponentially. Alternatively, with the ``Shifman
assumption'' \eqref{Shifmana} for $p$, which vanishes faster than
exponentially at large $N$,
one obtains instead from Eq.~\eqref{collapseb}:
\begin{equation}\lim_{N\rightarrow\infty}(\G_{kN})^{1/kN}\ =\
\Big({2e\over N}\Big)^{1/k}\,\J(N)\ .
\elabel{rootdefz}\end{equation}
Combining this with the large-$N$ limit of
the 1-instanton expression \eqref{SCIansb},
 one extracts instead the expression
\begin{equation}
{1\over16\pi^2}{\cal J}(N)=\ \Lambda^3\ 
\elabel{jnks}\end{equation}
which now agrees (by construction)  with the 1-instanton WCI calculation.

Below we will calculate $\G_{kN},$ to leading order in $1/N$, but for
all instanton number $k$, and will obtain a markedly different
behavior. Explicitly we will find:
\begin{equation}
{1\over16\pi^2}\lim_{N\rightarrow\infty}(\G_{kN})^{1/kN}\ =\ {2e\over
N}k\Lambda^3+\CO(N^{-2})\ .  \elabel{rootdefb}\end{equation} Notice
that for $k=1$ we obviously recover the results
\eqref{rootdef}-\eqref{jnclust} (or \eqref{rootdefz}-\eqref{jnks});
however the linear $k$ dependence is in sharp disagreement with the $k$
dependence of either Eq.~\eqref{rootdef} or Eq.~\eqref{rootdefz}.
This disagreement means that the generalized clustering assumption
\eqref{Pzerodef} is invalid when combined with the instanton
approximation. It also means that that the extension \eqref{Pzerodefa}
of this clustering assumption, in the presence of an extra KS vacuum
state, is likewise invalid.

The large-$N$ calculation proceeds as follows.\footnote{Our
conventions are taken from \cite{MO-III,KMS} which also provide
self-contained reviews of the ADHM formalism for the $SU(N)$ gauge
group.} In supersymmetric
theories, at topological level $k$, the bosonic and fermionic
collective coordinates live, respectively, in complex-valued matrices
$a$ and $\M,$ with elements:
\begin{equation}
a=\begin{pmatrix}w_{uj\dalpha}\\(a'_{\beta\dalpha})_{ij}^{}\end{pmatrix}\
,\qquad
\M=\begin{pmatrix}\mu_{uj}\\(\M'_\beta)_{ij}^{}\end{pmatrix}\ .
\elabel{adef}\end{equation}
The indices run over
\begin{equation}
u=1,\ldots,N\ ,\quad i,j=1,\ldots,k\ ,\quad
\dalpha,\beta=1,2\ ;
\elabel{indexrun}\end{equation}
traces over these indices are denoted `$\tr_N$', `$\tr_k$', and
`$\tr_2$', respectively. 
The elements of $\M$ are Grassmann (i.e., anticommuting)
quantities. The $k\times k$ submatrices $a'_{\beta\dalpha}\equiv
a'_n\sigma^n_{\beta\dalpha}$ and $\M'_\beta$ are subject to the
Hermiticity conditions
\begin{equation}\abar_n'=a_n'\ ,\quad\Mbar'_\alpha=\M'_\alpha\ .
\elabel{hermcond}\end{equation}
In the instanton approximation, the Feynman path integral is replaced
by a finite-dimensional integration over the degrees of freedom in $a$
and $\M.$ These $k$-instanton collective coordinates are weighted
according to the integration measure
\cite{meas1,meas2,MO-III,KMS}\footnote{The 
reason we have
$2^{k^2/2}$ rather than $2^{-k^2/2}$, as in \cite{MO-III}, is that we
restore Wess and Bagger integration conventions for the $\M'$
integration: $\int d^2\xi\,\xi^2=1$ rather than 2 where
$\xi^2=\xi^\alpha\xi_\alpha$ is the square of a Grassmann Weyl spinor.}
\def\dmuphys{d\mu^{k}_{\rm phys}}
\begin{equation}\begin{split}
\int\dmuphys\ &=\ {2^{k^2/2}(C_1)^k\over{\rm
Vol\,}U(k)}\int d^{4k^2}a'\,d^{2kN}\wbar\,d^{2kN}w\,d^{2k^2}\M'\,d^{kN}\mubar\,d^{kN}\mu
\\&\times\ 
\prod_{r=1,\ldots,k^2}\Big[\prod_{c=1,2,3}\delta\big(\hf{\rm tr}_k\,T^r(\trtwo\, \tau^c \abar a)\big)
\prod_{\dalpha=1,2}\delta\left({\rm tr}_k\,T^r(\Mbar a_\aD + \abar_\aD
\M)\right)\Big]\ ,
\elabel{dmuphysdef}\end{split}\end{equation}
where the two $\delta$-functions enforce the bosonic and fermionic
ADHM constraint conditions, respectively.  The integrals over the
$k\times k$ matrices $a'_n$ and $\M^{\prime}$ are defined as the
integral over the components with respect to a Hermitian basis of
$k\times k$ matrices $T^r$ normalized so that ${\rm
tr}_k\,T^rT^s=\delta^{rs}$. These matrices also provide explicit
definitions of the $\delta$-function factors in the way indicated.

The form of the measure given in Eq.~\eqref{dmuphysdef} is known as
the ``flat measure'', since the bosonic and fermionic ADHM collective
coordinates are integrated over as Cartesian variables, subject to the
nonlinear $\delta$-function constraints. It was uniquely constructed
in Ref.~\cite{meas1} to obey several important consistency
requirements---including cluster decomposition---so that the failure
of cluster uncovered below cannot be attributed to the collective
coordinate measure.  In practical applications, however, the flat
measure is not the most useful form available. When
\begin{equation}
N\ \ge\ 2k\ , \elabel{Nkinequality}\end{equation} it is convenient to
switch to the so-called ``gauge-invariant measure,'' involving a new
set of variables in terms of which the arguments of the
$\delta$-functions are linear (and hence trivially implemented)
\cite{MO-III}. This is the form of the measure which we will utilize
in the present section. The restriction \eqref{Nkinequality} is
obviously well suited to the large-$N$ limit. As the name implies, the
gauge-invariant measure can only be used to integrate gauge-invariant
quantities, such as our present focus on correlators formed from
$\trN\lambda^2.$ Alternatively, for the special cases $k\le2,$ it is
easy to solve the nonlinear constraints explicitly without such a
change of variables \cite{Osborn:1981yf,MO-I}.

In order to switch from the flat measure to the gauge-invariant
measure, one trades the collective coordinates $w$ and $\wbar$ (which
transform in the $N$ of $SU(N)$)  for the gauge-invariant
bosonic bilinear quantity $W$, defined by \cite{MO-III}
\begin{equation}
\big(W_{\ \dbeta}^\dalpha\big)_{ij}=\bar w_{iu}^\dalpha
 \,w_{uj\dbeta}\ ,\quad
W^0={\rm tr}_2\,W,\quad W^c={\rm
tr}_2\,\tau^cW, \ \ c=1,2,3\ .
\elabel{Wdef}\end{equation}
The appropriate Jacobian for this change of variables reads:
\begin{equation}
d^{2kN}\wbar\,d^{2kN}w\ \longrightarrow\
c_{k,N}\,\big(\det_{2k}W\big)^{N-2k}d^{k^2}
W^0\prod_{c=1,2,3}d^{k^2}W^c\ ,
\elabel{Jacis}\end{equation}
where
\begin{equation}
c_{k,N}\ =\
{2^{2kN-4k^2+k}\,\pi^{2kN-2k^2+k}\over\prod_{i=1}^{2k}(N-i)!}\ .
\elabel{ckNdef}\end{equation}
Note that the bosonic $\delta$-function in \eqref{dmuphysdef} can be
rewritten in a gauge-invariant way as the condition
\begin{equation}
0=W^c+[\,a'_n\,,\,a'_m\,]\,\trtwo\,\tau^c\sigmabar^{nm}= 
 W^c - i [\,a'_n\,,\,a'_m\,]\,\etabar^c_{nm} 
\elabel{adhmredux}\end{equation}
in terms of the gauge-invariant coordinates (here $\etabar^c_{nm}$ is
an `t Hooft tensor \cite{tHooft}). As advertised, these constraints are {\it
linear\/} in the new variables $W^c$; consequently the $W^c$ integrals simply remove
the bosonic ADHM $\delta$-functions in Eq.~\eqref{dmuphysdef} (giving rise to the
numerical factor of $2^{3k^2}$ from the $\hf$'s in the arguments of
the $\delta$-functions). 

Next we perform a similar change of variables for the fermions,
letting  \cite{MO-III}
\begin{equation}
\mu_{ui}=w_{uj\aD}(\zeta^{\dalpha
})_{ji}+\nu_{ui},\qquad
\bar\mu_{iu}=(\bar\zeta^{
}_\dalpha)_{ij}\bar w_{ju}^\dalpha+\bar\nu_{iu}\ ,
\elabel{zetadef}\end{equation}
where $\nu$ lies in the orthogonal subspace to $w$:
\begin{equation}
\bar w_{iu}^\aD\nu_{uj}=0\ ,\qquad \bar\nu_{iu} w^{}_{uj\aD}=0\ .
\elabel{nudef}\end{equation}
One finds:
\begin{equation}
\int d^{kN}\mu\,d^{kN}\mubar
\prod_{r=1,\ldots,k^2}\prod_{\dalpha=1,2}
\delta\left({\rm tr}_k\,T^r(\Mbar a_\aD + \abar_\aD \M)\right)
\ \longrightarrow
\ 2^{k^2}\int d^{2k^2}\zeta\,d^{kN-2k^2}\nu\,d^{kN-2k^2}\nubar\ ,
\elabel{fermiXn}\end{equation}
%\begin{equation}\fermiXn{\eqalign{&\int d^{kN}\mu\,d^{kN}\mubar
%\prod_{r=1,\ldots,k^2}\prod_{\dalpha=1,2}
%\delta\left({\rm tr}_k\,T^r(\Mbar a_\aD + \abar_\aD \M)\right)
%\\&\longrightarrow
%\ \int d^{2k^2}\zeta\,d^{kN-2k^2}\nu\,d^{kN-2k^2}\nubar}}
where the $\delta$-functions have been used to eliminate the
$\zetabar$ variables from the problem.
In summary, the gauge-invariant measure is:
\begin{equation}{2^{9k^2/2}\,(C_1)^k\,c_{k,N}\over{\rm Vol\,}U(k)}\int
d^{4k^2}a'\,d^{k^2}W^0\,d^{2k^2}\M'\,d^{2k^2}\zeta\,
d^{kN-2k^2}\nu\,d^{kN-2k^2}\nubar\,(\det_{2k}W)^{N-2k}
\elabel{GImeas}\end{equation}
and the constraint $\delta$-functions have been eliminated for all $k$
satisfying Eq.~\eqref{Nkinequality}.
For $k=1,$ one recovers the expression
\eqref{measone}, a comparison which fixes the
normalization constant $C_1$:
\begin{equation}
C_1\ =\ -2^{N+1/2}\Lambda^{3N}\ .
\elabel{Conedef}\end{equation}
For $k=2$, we recapture the Osborn measure
discussed in Refs.~\cite{Osborn:1981yf,MO-I,meas1}, which we utilize in
Sec.~VI below.

Into this measure we now insert
\begin{equation}
\trN\lambda^2(x_1)\times\cdots\times\trN\lambda^2(x_{kN})\ ,
\elabel{insertinto}\end{equation}
where the gluino $\lambda^\alpha(x)$ is replaced in the instanton
approximation by a general superposition of adjoint fermion zero
modes. In terms of the previously introduced collective coordinates
$a$ and $\M,$ a useful identity states \cite{MO-III}:
\begin{equation}
\trN\lambda^2(x)\ =\ -\quarter
\bbox\,\trk\Mbar(\P+1)\M f\ ,
\elabel{corriganid}\end{equation}
where the ADHM quantities $\P$ and $f$ are defined as:
\begin{equation}
\P=1-\Delta f\Deltabar\ ,\quad f=(\Deltabar\Delta)^{-1}\
,\quad\Delta=a+bx\ ,
\elabel{adhmdefs}\end{equation}
and $b$ is the $(N+2k)\times(2k)$ matrix whose lower $2k\times 2k$
part is the identity $\delta_\beta^{\ \alpha}\delta_{il}$ and whose
upper $N\times 2k$ part is zero (quaternionic multiplication is
implied in the product $bx$). As discussed earlier,
$\G_{kN}(x_1,\ldots,x_{kN})$ is actually a constant, independent of
the $x_i$. The $x_i$ can therefore be chosen for maximum simplicity of
the algebra. However, the simplest conceivable choice, $x_i=0$ for all $i$,
results in an ill-defined answer of the form ``$0\times\infty$'' (the
zero coming from unsaturated Grassmann integrations, and the infinity
from divergences in the bosonic integrations due to coincident poles);
we have already noted this fact in the 1-instanton sector in Sec.~II above.
The simplest choice of
the $x_i$ that avoids this problem turns out to be:
\begin{equation}\begin{split}
&x_1=\cdots=x_{kN-k^2}=0\
,\\&x_{kN-k^2+1}=\cdots=x_{kN}=x
\elabel{simple}\end{split}\end{equation}
which we adopt for the remainder of this section.\footnote{As a
nontrivial check on our algebra, we have also numerically integrated
the large-$N$ correlator for insertions other than Eq.~\eqref{simple}, and
verified the constancy of the answer presented below.}

In the large-$N$ limit the large preponderance of the insertions
\eqref{simple} are at $x_i=0,$ and the resulting factor of
$(\trN\lambda^2(0))^{kN-k^2},$ taken together with the Jacobian factor
$(\det_{2k}W)^{N-2k}$ from the measure \eqref{GImeas}, dominate the integral
and can be treated in
saddle-point approximation. Below we will carry out this saddle-point
evaluation in full detail, but we can already quite easily understand
the source of the linear dependence on $k$ in the final result
\eqref{rootdefb}. The chain of argument goes as follows:

\bf1\rm. Let us imagine carrying
out all the Grassmann integrations in the problem. The remaining
large-$N$ integrand will then have the form $\exp\big(-N\Gamma+\CO(\log
N)\big)$ where $\Gamma$ might be termed the ``effective large-$N$ bosonic
instanton action.'' The large-$N$ saddle-points are then the
stationary points of $\Gamma$ with respect to the bosonic collective
coordinates. By Lorentz symmetry, $\Gamma$ can only depend on
the four $k\times k$  matrices $a'_n$ through
even powers of $a'_n$. (This is because the bulk of the insertions
have been chosen to be
at $x_i=0$; otherwise one could form the Lorentz scalar $x^{}_na'_n$ and so
have odd powers of $a'_n$.) It follows that the ansatz
\begin{subequations}
\begin{align}
&a'_n=0\ ,\quad n=1,2,3,4\ ,\elabel{spansatza}\\&W^c=0\ ,\quad
c=1,2,3 \elabel{spansatzb}
\end{align}\end{subequations}
is automatically a stationary point of $\Gamma$ with respect to these
collective coordinates. (Note that \eqref{spansatzb} follows automatically
from \eqref{spansatza} by virtue of the ADHM constraints  
\eqref{adhmredux}.) It will actually turn out that, once one assumes these saddle-point
values, $\Gamma$ is independent of the remaining collective
coordinate matrix $W^0$; furthermore we will verify  that this
saddle-point is actually a  minimum of the Euclidean action.

\bf2\rm. Having anticipated the saddle-point
\eqref{spansatza}-\eqref{spansatzb} using these elementary
symmetry considerations, let us back up to a stage in the analysis
prior to the Grassmann integration, and proceed a little more carefully.
Evaluating the insertions $\trN\lambda^2(x_i)$ on this
saddle-point, one easily verifies that the $\zeta$ modes vanish when $x_i=0$;
consequently the $\zeta$ integrations must be saturated entirely from
the $k^2$ insertions at $x_i=x.$  This leaves the $\M',$ $\nu$ and
$\nubar$ integrations to be saturated purely from the insertions at
$x_i=0$. Moreover, because $\M'$ carries a Weyl spinor index $\alpha$ whereas
$\nu$ and $\nubar$ do not, the $\trN\lambda^2(0)$ insertions 
depend on these Grassmann coordinates only through  bilinears of the form
 $\nubar\times \nu$ or $\M'\times\M'$; there are no cross terms.

\bf3\rm. Performing all the Grassmann integrations then automatically generates
a combinatoric factor
\begin{equation}
(k^2)!\,(k^2)!\,(kN-2k^2)!\,\begin{pmatrix}kN-k^2\\ k^2\end{pmatrix}\ .
\elabel{combfact}\end{equation}
Here the first three factors account for the indistinguishable
bilinear insertions of the $\zeta,$ $\M',$ and $\{\nu,\nubar\}$ modes,
respectively, while the final factor counts the ways of
selecting the
$k^2$ bilinears in
 $\M'$ from  the $kN-k^2$ insertions at $x_i=0.$
Multiplying these combinatoric factors together, as well as the
normalization constants $c_{k,N}(C_1)^k$ from Eq.~\eqref{GImeas}, and taking the
$(kN)^{\rm th}$ root yields, in the large-$N$ limit:
\begin{equation}
\lim_{N\rightarrow\infty}\big[
c_{k,N}(C_1)^k\,(k^2)!\,(kN-k^2)!\big]^{1/kN}\ =\
2^{3}\pi^{2}eN^{-1}k\Lambda^3+\CO(N^{-2})\ .
\elabel{kNroot}\end{equation}
Remarkably, apart from a
factor of four, this back-of-the-envelope analysis precisely accounts
for the previously announced final answer, Eq.~\eqref{rootdefb}. Note that
most of the remaining contributions to the saddle-point analysis,
which involve a specific convergent bosonic integral derived below, as
well as the factor $2^{9k^2/2}/{\rm Vol\,}U(k)$ from Eq.~\eqref{GImeas}, reduce
to unity when the $(kN)^{\rm th}$ root is taken in the large-$N$
limit; the missing factor of four will simply come from the leading
saddle-point evaluation of the bosonic integrand.

Here are the details of the large-$N$ calculation of $\G_{kN}.$ 
Since the problem has an obvious $U(k)$ symmetry \cite{MO-III}, we will find it
convenient to work in a basis where $W^0$ (which transforms in the
adjoint of the $U(k)$) is diagonal:
\begin{equation}W^0\ =\
\begin{pmatrix}2\rho_1^2&{}&{0}\\{}&\ddots&{}\\{0}&{}&2\rho_k^2\end{pmatrix}\ .
\elabel{Wdiag}\end{equation}
As the notation implies, in the dilute instanton gas limit $\rho_i$
can be identified with the scale size of the $i^{\rm th}$ instanton in
the $k$-instanton sector (see Sec.~II.4 of \cite{MO-III}). The appropriate
change of variables reads:
\begin{equation}
{1\over{\rm Vol}\,U(k)}\int d^{k^2}W^0\ \longrightarrow\ {2^{3k(k-1)/2}\pi^{-k}\over k!}
\int_0^\infty
d\rho_1^2\cdots d\rho_k^2\,\prod_{1\le i<j\le k}(\rho_i^2-\rho_j^2)^2\ .
\elabel{diagcov}\end{equation}
%where the normalization constant $\A_k$ is 
%fixed  by integrating both
%sides against the convergence factor $\exp(-\trk(W^0)^2)\,$:
%\begin{equation}
%\A_k\ =\  {2^{3k(k-1)/2}\pi^{-k}\over k!}\ .
%2^{-k}\pi^{k^2/2}\Big(
%\int_0^\infty
%d\rho_1^2\cdots d\rho_k^2\,\exp(-4\sum_{i=1}^k\rho_i^4)\,
%\prod_{1\le i<j\le k}(\rho_i^2-\rho_j^2)^2\,\Big)^{-1}\ .
%\elabel{Akdef}\end{equation}
%The factor of $2^{-k}$ comes from the positivity constraints on the
%eigenvalues of $W^0$. 
For $k=1$ one has, of course, $\int
dW^0\rightarrow 2\int_0^\infty d\rho^2.$ 

Now let us consider the Grassmann integrations, beginning with the
$\zeta$ modes. We assume the saddle-point conditions \eqref{spansatza}-\eqref{spansatzb}, in
which case
\begin{equation}
\Delta=\begin{pmatrix}w\\ x\cdot1_{\sst [k]\times[k]}^{}\end{pmatrix}\ ,\qquad
f=\begin{pmatrix}{1\over\rho_1^2+x^2}&{}&{0}\\ {}&\ddots&{}\\{0}&{}&
{1\over\rho_k^2+x^2}\end{pmatrix}\ ,
\elabel{spmore}\end{equation}
and from Eq.~\eqref{corriganid},
\begin{equation}
\trN\lambda^2(x)\ =\ -\sum_{i,j=1}^k(\zeta_\dalpha)_{ij}
(\zeta^\dalpha)_{ji}\,F_{ij}(x)\ +\ \cdots\ ,
\elabel{zetastuff}\end{equation}
where
\begin{equation}
F_{ij}(x)\ =\
\quarter\bbox\,{x^4\over(x^2+\rho_i^2)(x^2+\rho_j^2)}
\elabel{Fijdef}\end{equation}
and the omitted terms in Eq.~\eqref{zetastuff} represent dependence on the
other Grassmann modes $\{\M',\nu,\nubar\}$. It is obvious from
Eq.~\eqref{Fijdef} that $F_{ij}(0)=0$, so that the $\zeta$ modes must be
entirely saturated from the $k^2$ insertions at $x_i=x$ as claimed
above. Performing the $\zeta$ integrations then yields
\begin{equation}(-1)^{k^2}(k^2)!\,\prod_{i,j=1}^kF_{ij}(x)\ .
\elabel{zetaans}\end{equation}

Next we consider the insertions at $x_i=0$. Focusing on the $\M'$ modes
first, one finds from Eq.~\eqref{corriganid}:
\begin{equation}
\trN\lambda^2(0)\ =\
2\sum_{i,j=1}^k(\M^{\prime\alpha})_{ij}(\M'_{\alpha})_{ji}\,
(\rho_i^{-4}+\rho_j^{-4}+\rho_i^{-2}\rho_j^{-2})\
 +\ \cdots\ ,
\elabel{Mpstuff}\end{equation}
omitting the $\nu\times\nubar$ terms. Hence the $\M'$ integrations yield
\begin{equation}
\begin{pmatrix}kN-k^2\\ k^2\end{pmatrix}\,(k^2)!\,2^{k^2}\,\prod_{i,j=1}^k
(\rho_i^{-4}+\rho_j^{-4}+\rho_i^{-2}\rho_j^{-2})\ ,
\elabel{Mpans}\end{equation}
where the combinatoric factors in \eqref{Mpans} (as well as in \eqref{zetaans})
have been explained previously.\footnote{One can easily check that
these large-$N$ formulae are consistent with the explicit 1-instanton
calculation presented in Sec.~II which is exact in $N$. In particular,
if one takes $x_i=x$ while $x_j=0,$ then
Eqs.~\eqref{threeconb}-\eqref{threeconc} are suppressed
vis-$\grave{\rm a}$-vis Eq.~\eqref{threecona} by factors of $a'_n$, and
in turn, $a'_n\sim N^{-1/2}$ as follows from Eq.~\eqref{nuans} below.}

Finally we turn to the $\{\nu,\nubar\}$ integrations. Since (unlike
the $\zeta$ and $\M'$ modes) the number of $\nu$ and $\nubar$ modes
grows with $N$ as $kN-2k^2,$ it does not suffice merely to plug in the
saddle-point values \eqref{spansatza}-\eqref{spansatzb} and \eqref{spmore}. One must also calculate
the Gaussian determinant about the saddle-point, which provides an
$\CO(N^0)$ multiplicative contribution to the answer. Accordingly we
expand about \eqref{spansatza}-\eqref{spansatzb} 
to quadratic order in the $a'_n$. The $\nu\times\nubar$
contribution to $\trN\lambda^2(0)$ has the form
\begin{equation}
-\hf\nubar_{ju}\nu_{ui}\bbox f_{ij}{\Big|}_{x=0}\ =\ 
2\nubar_{ju}\nu_{ui}\big(f\cdot\trtwo \bbar\P b\cdot f\big)_{ij}
{\Big|}_{x=0}
\elabel{nunubar}\end{equation}
as follows from Eqs.~\eqref{corriganid}-\eqref{adhmdefs}, and Eq.~(2.63)
of \cite{MO-III}.
Performing
the $\{\nu,\nubar\}$ integrations therefore gives
\begin{equation}\begin{split}
&(kN-2k^2)!\,\exp\Big((N-2k)\trk\log\big(2f\cdot\trtwo \bbar
\P b\cdot
f\big){\Big|}_{x=0}\,\Big)\ =
\\ &
(kN-2k^2)!\,\exp\Big((N-2k)\big(\log\det_k16(W^0)^{-2}\, -\,
\tfrac32\sum_{i,j=1}^k \sum_{n=1}^4a'_{nij}a'_{nji}\,(\rho_i^{-2}+\rho_j^{-2})\,
+\, \CO(a'_n)^4\big)\,\Big)\ .
\elabel{nuans}\end{split}\end{equation}
The negative sign in front of the quadratic term in $a'_n$ confirms
that our saddle-point \eqref{spansatza}-\eqref{spansatzb} is in fact a minimum of the action.
Combining this expression with the measure factor in Eq.~\eqref{GImeas}, namely
\begin{equation}\begin{split}
(\det_{2k}W)^{N-2k}\ &=\ \exp\big((N-2k)\log\det_{2k}W\big)
\\&=\
\exp\Big((N-2k)\big(\log\det_k(\tfrac12W^0)^{2}\,+\,\CO(a'_n)^4\,
\big)\Big)\ ,
\elabel{measfact}\end{split}\end{equation}
and performing the Gaussian integrations over $a'_n$, yields:
\begin{equation}
2^{2k(N-2k)}\,(kN-2k^2)!\,\prod_{i,j=1}^k
\Big({2\pi\over3N(\rho_i^{-2}+\rho_j^{-2})}\Big)^2
\ +\ \cdots\ ,
\elabel{nuansb}\end{equation}
where the omitted terms are suppressed by powers of $N$.

Finally one combines Eqs.~\eqref{GImeas}, \eqref{diagcov}, \eqref{zetaans}, \eqref{Mpans} and
\eqref{nuansb} to obtain the leading-order result for the
correlator:
\begin{equation}
\lim_{N\rightarrow\infty}\G_{kN}\ =\
{2^{5kN+k^2-k+1/2}\,\pi^{2kN-k+1/2}\,e^{kN}\,(k^2)!\,k^{kN-k^2+1/2}\,{\cal
I}_k\,\Lambda^{3kN}\over3^{2k^2}\,N^{kN+k^2-1/2}\,k!}\ ,
\elabel{Gnfinal}\end{equation}
where $\I_k$ is the convergent integral
\begin{equation}
\I_k\ =\ 
\int_0^\infty
d\rho_1^2\cdots d\rho_k^2\,\prod_{1\le i<j\le
k}(\rho_i^2-\rho_j^2)^2\cdot\prod_{i,j=1}^k
F_{ij}(x)\,\big(1-(\rho_j/\rho_i+
\rho_i/\rho_j)^{-2}\big)\ .
\elabel{Ikdef}\end{equation}
Note that $\I_k$ is independent of $x$ as a simple rescaling argument confirms.
For the simple case $k=1$, the $(\rho_i^2-\rho^2_j)^2$ terms in this
integral are absent; one finds
$\I_1=\tfrac32$ and the expression \eqref{Gnfinal} 
agrees---as it must---with the large-$N$ limit of the 1-instanton SCI
result \eqref{finalone}.

\rsen
\section{The 4-Point Function of the Gluino Condensate in $SU(2)$ Gauge Theory}

We have seen that cluster decomposition fails (both with and without a
KS vacuum) in the SCI calculation of the gluino condensate, for gauge
group $SU(N)$ in the large-$N$ limit. In this section we focus instead
on the gauge group $SU(2)$. In this case, at the 1-instanton level,
the 2-point function \eqref{SCIans} works out to:
\begin{equation}
\VEV{{\tr\lambda^2(x_1)\over16\pi^2}\,{\tr\lambda^2(x_2)\over16\pi^2}}\
=\ \textstyle{4\over5}\,\Lambda^{6}\ .
\elabel{SCItwo}\end{equation}
Here we will calculate the 4-point function, which receives a nonzero
contribution at the 2-instanton level:
\begin{equation}
\VEV{{\tr\lambda^2(x_1)\over16\pi^2}\ {\tr\lambda^2(x_2)\over16\pi^2}\ 
{\tr\lambda^2(x_3)\over16\pi^2} \ 
{\tr\lambda^2(x_4)\over16\pi^2}}\
=\ c\Lambda^{12}\ .
\elabel{SCIfour}\end{equation}
In the absence of a KS vacuum, generalized cluster decomposition
together with Eq.~\eqref{SCItwo}
predicts $c=(4/5)^2=.64$. Alternatively, in the presence of a KS vacuum,
weighted according to Eq.~\eqref{Shifmana} in order to reconcile the SCI and
WCI 1-instanton calculations, one expects $c=4/5=.8$. Instead, we have
calculated $c$ numerically, and find:
\begin{equation}
c\ \simeq\ .500\pm.026\ .
\elabel{cdef}\end{equation}
Here are the details of the calculation.

As mentioned above, for $k=2,$ one can eliminate the $\delta$-function
constraints in Eq.~\eqref{dmuphysdef} without changing variables. Another
simplification for the particular gauge group $SU(2)$ is that one can
adopt a concise quaternionic representation for the ADHM bosonic
collective coordinates, taking advantage of the fact that $SU(2)\cong
Sp(1).$ Specifically,
the $16$ gauge and $8$ gaugino 
 collective coordinates live, respectively, in the following
matrices:\footnote{See Ref.~\cite{MO-I} for details of notation and
conventions pertinent to Sec.~VI.}
\begin{equation}
a\ =\ \begin{pmatrix}w_1&w_2\\ a'_{11}&a'_{12}\\ a'_{12}&a'_{22}\end{pmatrix}\ ,\qquad
\M_\gamma\ =\ \begin{pmatrix}\mu_{1\gamma}&\mu_{2\gamma}\\
\M'_{11\gamma}&\M'_{12\gamma}\\
\M'_{12\gamma}&\M'_{22\gamma}\end{pmatrix}\ ,
\elabel{twoinstmats}\end{equation}
where $a=a_{\alpha\dalpha}=a_n\sigma^n_{\alpha\dalpha}$ and the
matrices $a_n$ as well as $\M_\gamma$ are real-valued (unlike the
complex-valued collective coordinates of the same name introduced in
Eq.~\eqref{adef} which are needed for general $SU(N)$).  The resulting
2-instanton ``Osborn measure'' on these collective coordinates is
detailed in Refs.~\cite{Osborn:1981yf,MO-I,meas1}, and reads:
\def\dmuphystwo{d\mu^{2}_{\rm phys}}
\begin{equation}
\int\dmuphystwo\ =\ 2^{14}\Lambda^6\int
d^4w_1\,d^4w_2\,d^4a'_{11}\,d^4a'_{22}\, d^2\mu_1\,d^2\mu_2\,d^2\M'_{11}\,
d^2\M'_{22}\ {\big|\,|a'_3|^2-|a'_{12}|^2\,\big|\over|a'_3|^2}\ .
\elabel{osbdef}\end{equation}
%\begin{equation}\osbdef{\eqalign{\int\dmuphystwo\ =\ 2^{14}\Lambda^6\int&
%d^4w_1\,d^4w_2\,d^4a'_{11}\,d^4a'_{22}\\&\times\ d^2\mu_1\,d^2\mu_2\,d^2\M'_{11}\,
%d^2\M'_{22}\ {\big|\,|a_3|^2-|a'_{12}|^2\,\big|\over|a_3|^2}}}
Here the $\delta$-function constraints from the flat measure have been
used to eliminate $a'_{12}$ and $\M'_{12}$ in terms of the other
collective coordinates, via:
\begin{equation}
a'_{12}\ =\
{1\over4|a'_3|^2}\,a'_3(\wbar_2w_1-\wbar_1w_2)\ ,
\elabel{aonedef}\end{equation}
and
\begin{equation}
\M'_{12}\ =\ {1\over2|a'_3|^2}\,a'_3\,\big(\,2\abar'_{12}\M'_3+\wbar_2\mu_1
-\wbar_1\mu_2\,\big)\ ,
\elabel{Monedef}\end{equation}
where we have defined
\begin{equation}
a'_3\,=\,\hf(a'_{11}-a'_{22})\
,\qquad\M'_3\,=\,\hf(\M'_{11}-\M'_{22})\ .
\elabel{athreedef}\end{equation}
Into this measure one inserts the 4-point function of the classical condensate,
expressed as a function of the 2-instanton collective coordinates \eqref{twoinstmats}. 
The 8-dimensional Grassmann integrations over $\{ \mu_1, \mu_2, 
{\cal M}'_{11}, {\cal M}'_{22} \} $ are then accomplished in two steps.
The first step is to expand the integrand in terms of Grassmann variables
using a modified version of the program ``Dill'',
written for {\scshape Mathematica}.\footnote{``Dill'' is 
a {\scshape Mathematica} package originally written by Vladan Lucic
\cite{lucic} in 1994 
in order to simplify SUSY algebraic expressions. 
This program can be modified so that it can  
handle the large number of Grassmann variables that we need.}
The second step involves the explicit Grassmann integration, accomplished
using an ``awk-script'' implemented on a UNIX system and 
made to perform the symbolic algebra of 
Grassmann integration.

The resulting 16-dimensional bosonic integration over $ \{ w_1, w_2,
a'_{11}, a'_{22} \} $, the remaining quaternionic 
variables, is carried out
using a standard Monte Carlo integration procedure.
The integrable singularities are handled using the standard procedure:
firstly, dropping a tiny region around the integrable singularities and
then making sure that the contribution from this dropped region is 
negligibly smaller than the precision required.
After 450 million points have been sampled, we have obtained the numerical value
\eqref{cdef} given above. As a check on our numerics, we have also verified
the constancy of the answer by comparing different choices for the
four space-time insertion points.

\rsen
\section{Discussion}

The mismatch between the strong coupling and weak coupling
calculations is a fascinating puzzle. Previously, only the mismatch at
the one instanton level was known; now we see a mismatch established
at large $N$ for all instanton numbers, and for $N=2$ at the
2-instanton level. Certainly we do not mean to imply that, because of
this mismatch, SCI calculations are all necessarily suspect; indeed an
$\N=4$ supersymmetric version of an SCI calculation performed by some
of us \cite{MO-III} has recently provided a dramatic quantitative and
qualitative verification of Maldacena's conjecture. However in this
case the coupling does not run and the calculation can be performed at
weak coupling, actually small $g^2N$, 
where the instanton approximation is fully justified. The
continuation to strong coupling, large $g^2N$, is then accomplished by means of a
non-renormalization theorem. Rather, our
objections to the SCI computation 
are more narrow and technical in scope: specifically, our
calculations imply a fundamental breakdown of clustering in the
instanton approximation to the gluino condensate at strong coupling.

One may wonder what the origin for this breakdown is? The usual
justification for the strong coupling calculation is that one can take
$|x_i-x_j|$ much smaller than the scale of strong coupling effects
$\Lambda^{-1}$ and so the theory would be weakly coupled, due to
asymptotic freedom, and the instanton calculation would be
justified. Then, since the correlation functions \eqref{Gndef} are
independent of the positions, the result would be valid at all
distances.  This point-of-view has simultaneously been used and
criticized by various authors \cite{Rossi:1984bu,
Novikov:1985ic,Shifman:1988ia}.  The asymptotic freedom argument means
that the first-order, second-order, etc., perturbative corrections to
the SCI calculations are small---indeed, for the gluino condensate
correlators discussed herein, these perturbative corrections are
entirely absent due to a nonrenormalization theorem
\cite{Novikov:1985ic}. However,
the asymptotic freedom argument does not guarantee that the zeroth
order instanton calculation is itself complete; there may be
other non-perturbative configurations contributing to the
correlators. Indeed, the breakdown of cluster suggests that such
additional nonperturbative configurations (with
size of order $\Lambda^{-1}$) must be present, and that they must
account for the mismatch between the SCI and WCI calculations.

In contrast, it seems that the WCI
calculation uses a method that has amassed a considerable
pedigree. These kinds of calculations appear to be consistent in all
applications and agree with other non-instanton methods
\cite{Finnell:1995dr}; for example, the two-instanton check of the
Seiberg-Witten approach to ${\cal N}=2$ theories \cite{MO-I,MO-II} and
the latter calculation in Sec.~III. Moreover, in the WCI set-up,
large-scale nonperturbative configurations as just discussed, such as
instanton-antiinstanton pairs of size $\Lambda^{-1}$, would be
exponentially suppressed in the path integral so long as
$\Lambda\ll\rmv$. We should further note that, as the separation
between insertions tends to zero, the WCI calculation does \it not \rm
smoothly go over to the SCI calculation as one would naively expect; there are additional
important contributions which will be discussed in a separate publication (work
in progress). 

%It seems to us that, in the strong coupling
%calculation, we are doing a
%semi-classical approximation, namely the instanton approximation,
%outside the range of its validity. There is no reason to suppose that 
%other non-perturbative, but non-instantonic, configurations
%will not contribute to the correlation functions and reconcile the
%strong and the weak coupling calculations. 
It is unfortunate that, based on our results, 
the highly original and intriguing (both theoretically and phenomenologically)
proposal of Shifman, namely the existence of a chirally
symmetric vacuum state, loses much of its {\it raison d'\^etre\/}. Then
again, we have not actually ruled out the existence of such a state.
After the completion of this work, it has been suggested that the
mixing parameter $p$ of the KS vacuum, defined in
Eq.~\eqref{varrhopdef} above,  may actually be instanton number
dependent \cite{shifpc}. {\it Prima facie\/}, this appears to be
incompatible with invariance under large gauge transformations 
($|k\rangle\rightarrow|k+1\rangle$);
however, if such a counter-intuitive flexibility is
permissible in the definition of the instanton vacuum, 
clustering in the presence of the KS vacuum can be saved.

\def\Ibar{\bar{\rm I}}
Conceptual difficulties with the instanton approximation and cluster
decomposition were pointed out in the context of 
pure (non-supersymmetric) QCD some time ago. Since this may have some
bearing on the present discussion, we review some comments of L\"uscher
regarding this issue \cite{Luscher:1979yd}. The
pure instanton (i.e.~no anti-instantons) approximation to QCD
obviously violates parity since
\begin{equation}
\VEV{\tr F_{nm}{}\,^*F_{nm}}_{\rm inst.}=\rho\neq0\ ,
\end{equation}
where $\rho$ here is the instanton density.
Parity is then recovered by summing over instantons (I) {\it and\/}
anti-instantons ($\Ibar$); however, in this approximation the cluster property
would not hold. To see this note that
\begin{equation}
\VEV{\tr F_{nm}{}\,^*F_{nm}(x)\ \tr F_{pq}{}\,^*F_{pq}(0)}\
\underset{|x|\rightarrow\infty}\rightarrow \ \rho^2\neq0\ ,
\label{sdfds}
\end{equation}
whereas 
\begin{equation}
\VEV{\tr F_{nm}{}\,^*F_{nm}}=\tfrac12\big(\VEV{\tr F_{nm}{}\,^*F_{nm}}_{\rm
inst.}+\VEV{\tr F_{nm}{}\,^*F_{nm}}_{\rm anti\hbox{\rm-}inst.}\big)=0\ .
\end{equation}
In order to resolve this clustering conundrum, it is apparent that
additional configurations, which may, in the dilute gas limit, be
thought of as mixtures of
instantons and anti-instantons, would need to be incorporated in the
approximation. In this case the two-point function \eqref{sdfds} would
indeed be zero, the result of summing the I$\,$I, I$\,\Ibar,$ $\Ibar\,$I and
$\Ibar\,\Ibar$ contributions which on average all contribute equally. Away from the dilute instanton gas limit, the identification and physical interpretation of these additional cluster-restoring nonperturbative configurations is necessarily more subtle.

In summary, the results of this paper imply something analogous
in the ${\cal N}=1$ theory: additional
configurations must contribute to the correlators
at strong coupling and resolve the breakdown of clustering (as well as
repairing the mismatch between the SCI and WCI calculations). 
In fact, it was suspected some time ago 
(see Ref.~\cite{Osborn:1981yf,Belavin:1979fb} and references therein) 
that in strongly coupled theories, it may be more appropriate to think
of instantons as composite configurations of some more basic objects:
so-called ``instanton partons''. 
The dominant contributions to the path integral at
strong coupling would then arise from the partons themselves.
In Ref.~\cite{DHKM}, we make this piece of folklore more precise by 
identifying instanton partons with the monopole configurations of the 
supersymmetric Yang-Mills compactified on the cylinder
${\Bbb R}^3 \times S^1$, with the circle having circumference $\beta$.
Each monopole has precisely {\it two\/} gluino zero modes, rather than
four for the instanton. The instanton itself is then identified with a 
specific two-monopole
configuration. We calculate the monopole contribution to the gluino
condensate and then, at the end of the day, 
take the decompactification limit $\beta \to \infty$.
The value of the gluino condensate obtained in this way,
is precisely the WCI result \eqref{WCIans}.

\section*{Acknowledgments}

We thank Philippe Pouliot, Larry Yaffe, Alex Kovner, Misha Shifman and
Nick Dorey for valuable discussions.
VVK and MPM acknowledge a NATO Collaborative Research Grant.
TJH and VVK acknowledge the TMR network grant FMRX-CT96-0012.

\startappendix

\rsen
\Appendix{Constancy of the Correlators}

Since the constancy of the correlation function \eqref{Gndef} plays such an
important role in our analysis, in this appendix, we review the
arguments leading to this result. More importantly, we explain how the
field theory proof remains valid in the instanton approximation. 

First the field theory proof \cite{Novikov:1983ee,Rossi:1984bu}. 
The argument is completely general and
applies to the correlation functions of any {\it lowest component\/} $A$ of a
gauge invariant  chiral superfield $\Phi$:
\begin{equation}
\Phi=A(x)+\sqrt2\theta\psi(x)+\cdots\ .
\end{equation} 
In the present discussion, the operator
$\tr\,\lambda^2$ is the lowest component of the
chiral superfield $\tr\, W^\alpha W_\alpha$, where $W_\alpha$ is
supersymmetric field strength.
Consider the correlation function
\begin{equation}
\VEV{A_1(x_1)\cdots A_p(x_p)}\ .
\end{equation}
We will show that this is independent of the $x_i$'s. To this end, one
has
\begin{equation}
{\partial\over\partial x^n}A_i(x)={i\over4}\bar\sigma_n^{\aD\alpha}
\{\bar Q_\aD,\psi_{i\alpha}(x)\}\ .
\elabel{derivact}\end{equation}
Hence
\begin{equation}\begin{split}
{\partial\over\partial x^n_i}&\VEV{A_1(x_1)\cdots A_p(x_p)}\\
=&\tfrac i4\bar\sigma_n^{\aD\alpha}\langle0\vert A_1(x_1)\cdots A_{i-1}(x_{i-1})
\{\bar Q_\aD,\psi_{i\alpha}(x_i)\}A_{i+1}(x_{i+1})\cdots A_p(x_p)\vert0\rangle\\
=&-\tfrac i4\bar\sigma_n^{\aD\alpha}\sum_{j=1}^{i-1}\langle0\vert 
A_1(x_1)\cdots[\bar Q_\aD,A_j(x_j)]\cdots
A_{i-1}(x_{i-1})\psi_{i\alpha}(x_i)A_{i+1}(x_{i+1})\cdots A_{p}(x_p)\vert0\rangle\\
+&\tfrac
i4\bar\sigma_n^{\aD\alpha}\sum_{j=i+1}^{p}\langle0\vert 
A_{1}(x_1)\cdots A_{i-1}(x_{i-1})
\psi_{i\alpha}(x_i)A_{i+1}(x_{i+1})\cdots [\bar
Q_\aD,A_{j}(x_j)]
\cdots A_{p}(x_p)\vert0\rangle\ ,
\elabel{commut}\end{split}\end{equation}
where the last line follows by commuting the $\bar Q_\aD$ through the
other insertions, to the left and right, respectively, until it hits
the vacuum which it annihilates. But $[\bar Q_\aD,A_{j}(x)]=0$ and
therefore the right-hand side of \eqref{commut} vanishes and
consequently the correlation function is, indeed, independent of the
insertion points. {\it QED\/}.

If the multiple correlator of $\tr\,\lambda^2$ is constant in the full
field theory, it then becomes an issue as to whether this constancy is
retained in the instanton approximation. That it is, rests upon two
facts. Firstly, the supersymmetry transformations of the fields can be
traded for supersymmetry transformations of the collective coordinates
\cite{Novikov:1983ee,meas1,meas2,KMS,MO-II}. In other words, the
supersymmetry algebra is represented on the collective
coordinates. Specifically, under an infinitesimal supersymmetry
transformation $\xi Q+\bar\xi\bar Q$:
\begin{subequations}
\begin{align}
&\delta a_\aD=\bar\xi_\aD\M,\qquad \delta\bar
a^\aD=-\bar\M\bar\xi^\aD\elabel{susyta}\\
&\delta\M=-4ib^\alpha\xi_\alpha,\qquad \delta\bar\M=4i\xi^\alpha\bar
b_\alpha\ .\elabel{susytb}
\end{align}
\end{subequations}
In particular, \eqref{derivact} will hold, with the fields replaced
by their expression in the instanton background and with the right-hand side
involving the appropriate transformation of the collective
coordinates. {\it Ipso facto\/}, the argument leading to \eqref{commut} will hold 
with the transformations acting on the collective coordinates;
moreover $[\bar Q_\aD,A]=0$, understood as a transformation of the
collective coordinates. The remaining piece of the proof is the
analogue of the fact that $\bar Q_\aD$ annihilates the vacuum state. In
the instanton approximation, where the functional integral is
approximated by the integral over the collective coordinates, the
analogue of the 
statement that the vacuum is a supersymmetry invariant, is the
statement that the measure on the space of collective coordinates is
invariant under the supersymmetry transformations \eqref{susyta}-\eqref{susytb}. 
This invariance was proved in \cite{meas1,meas2}.

\end{document}